SPECIAL RELATIVITY IN QUANTUM PHASE SPACE

D. Dragoman – Univ. Bucharest, Physics Dept., P.O. Box MG-11, 077125 Bucharest, Romania, e-mail: danieladragoman@yahoo.com

ABSTRACT:

A phase space treatment of special relativity of quantum systems is developed. In this approach a quantum particle remains localized if subject to inertial transformations, the localization occurring in a finite phase space area. Unlike non-relativistic transformations, relativistic transformations generally distort the phase space distribution function, being equivalent to aberrations in optics.

1. INTRODUCTION

Although classical relativity, and particularly special relativity, is a well-established theory, no agreement about its quantum counterpart exists. Controversies are raised even by the possibility of unifying quantum mechanics and relativity, which are based on quite different principles. However, quantum mechanics and classical physics, in particular classical optics, share the same mathematical tool if expressed in the phase space (PS) formalism: the Wigner distribution function [1-2]. The quest for a quantum relativity theory would benefit from a PS formulation, since then classical and quantum aspects could be more easily compared. Such a PS formulation of special relativity is the object of this paper. No such study has been undertaken up to now.

It is fair to say that a PS description of special relativity transformations has been already attempted, [3] based on the observation that the effect of a Lorentz boost on the coordinates in a four-dimensional Minkowski space is similar to PS squeezing in light-cone coordinates. Therefore, the results obtained using this approach are just an illustration of the equivalence of boosts with light-cone coordinate squeezing and do not actually constitute a PS approach to special relativity. On the contrary, a PS quantum relativity theory is put forward in this paper starting from quite different principles: the Poincare group approach to special relativity is reformulated in the PS formalism. The advantages of this standpoint are mainly based on the PS constancy of the area occupied by the quantum particle, which overcomes the criticisms regarding the spreading of the quantum wavefunction in the position representation and hence the impossibility of quantum localization, and on the simple geometrical significance of the Lorentz transformations when viewed in PS. The PS approach to special relativity is particularly useful in the conceptual simplification of the classical-quantum correspondence in relativistic theory.

## 2. THE POINCARE GROUP OF RELATIVISTIC QUANTUM PARTICLES

The principle of relativity states that the laws of physics are the same in frames of references/laboratories $F$ related through an inertial transformation such as: translation in time with $t$, translation in space with $\boldsymbol{r}=(x,y,z)$, boost characterized by velocity $\boldsymbol{v}=(v_x,v_y,v_z)$, or rotation by vector $\boldsymbol{\varphi}=(\phi_x,\phi_y,\phi_z)$. These inertial transformations form the Poincare group, their corresponding generators: the Hamiltonian $H$, defined as $\partial/\partial t$, the momentum $\boldsymbol{P}=(\partial/\partial x,\partial/\partial y,\partial/\partial z)$, the boost $\boldsymbol{K}=(\partial/\partial v_x,\partial/\partial v_y,\partial/\partial v_z)$, and the angular momentum $\boldsymbol{J}=(\partial/\partial\phi_x,\partial/\partial\phi_y,\partial/\partial\phi_z)$, constituting the basis of the associated Lie algebra. These generators do not commute, as can be seen from the following relations [4]:

$$[J_i,P_j]=J_iP_j-P_jJ_i=\varepsilon_{ijk}P_k,\quad [J_i,J_j]=\varepsilon_{ijk}J_k,\quad [J_i,K_j]=\varepsilon_{ijk}K_k,$$
$$[J_i,H]=[P_i,P_j]=[P_i,H]=0, \tag{1}$$
$$[K_i,K_j]=-c^{-2}\varepsilon_{ijk}J_k,\quad [K_i,P_j]=-c^{-2}H\delta_{ij},\quad [K_i,H]=-P_i,$$

where $i,j,k=x,y,z$, $\varepsilon_{ijk}$ is the anti-symmetric Levi-Civita symbol, $\delta_{ij}$ is the Kronecker delta function, $c$ is the vacuum light velocity, and the summation over repeated indices is assumed. The boost parameter $\boldsymbol{v}$ is customarily replaced by $c\boldsymbol{\theta}$, where the rapidity $\theta=|\boldsymbol{\theta}|$ is related to $\boldsymbol{v}$ through $\cosh\theta=(1-v^2/c^2)^{-1/2}$, or $\boldsymbol{v}(\boldsymbol{\theta})=(\boldsymbol{\theta}/\theta)c\tanh\theta$.

A general inertial transformation of a reference frame $F$ can then be expressed as

$$F'=T(\boldsymbol{\varphi},\boldsymbol{v},\boldsymbol{r},t)F=\exp(\boldsymbol{J}\boldsymbol{\varphi})\exp(\boldsymbol{K}c\boldsymbol{\theta})\exp(\boldsymbol{P}\boldsymbol{r})\exp(Ht)F. \tag{2}$$

The order of transformations in the above formula is assumed throughout the paper; the ordering is important since the transformations do not commute.

The group approach to the relativistic transformation of quantum states is based on the unitary representation of the Poincare group in the Hilbert space developed by Wigner [5]. More precisely, inertial transformations $T$ between reference frames $F$ and $F'=TF$ lead (in the Schrödinger representation) to transformations of quantum state vectors from $|\Psi\rangle$ to $|\Psi'\rangle = U_T |\Psi\rangle$, where

$$U_T = \exp(-i\hat{\boldsymbol{J}}\boldsymbol{\varphi}/\hbar)\exp(-i\hat{\boldsymbol{K}}c\boldsymbol{\theta}/\hbar)\exp(-i\hat{\boldsymbol{P}}\boldsymbol{r}/\hbar)\exp(i\hat{H}t/\hbar) \quad (3)$$

is the corresponding unitary operator in the Hilbert space [4]. The Hermitian operators of Hamiltonian $\hat{H}$, momentum $\hat{\boldsymbol{P}}$, angular momentum $\hat{\boldsymbol{J}}$, and boost $\hat{\boldsymbol{K}}$ are the generators of the Poincare group in the Hilbert space and satisfy the following commutation rules:

$$[\hat{J}_i, \hat{P}_j] = i\hbar\varepsilon_{ijk}\hat{P}_k, \quad [\hat{J}_i, \hat{J}_j] = i\hbar\varepsilon_{ijk}\hat{J}_k, \quad [\hat{J}_i, \hat{K}_j] = i\hbar\varepsilon_{ijk}\hat{K}_k,$$
$$[\hat{J}_i, \hat{H}] = [\hat{P}_i, \hat{P}_j] = [\hat{P}_i, \hat{H}] = 0, \quad (4)$$
$$[\hat{K}_i, \hat{K}_j] = -i\hbar c^{-2}\varepsilon_{ijk}\hat{J}_k, \quad [\hat{K}_i, \hat{P}_j] = -i\hbar c^{-2}\hat{H}\delta_{ij}, \quad [\hat{K}_i, \hat{H}] = -i\hbar\hat{P}_i,$$

In the Heisenberg representation of quantum mechanics, on the contrary, the state vectors remain constant, while observables transform as $\hat{O}' = U_T \hat{O} U_T^{-1}$, the expectation values of observables remaining the same (in agreement with the relativity principle) when both system and observables are subject to inertial transformations: $\langle\hat{O}'\rangle = \langle\Psi'|\hat{O}'|\Psi'\rangle = \langle\Psi|\hat{O}|\Psi\rangle = \langle\hat{O}\rangle$.

In terms of the Poincare group generators one can define the velocity operator as $\hat{\boldsymbol{V}} = \hat{\boldsymbol{P}}c^2/\hat{H}$ and two invariant Casimir operators: the mass $\hat{M} = c^{-2}(\hat{H}^2 - \hat{\boldsymbol{P}}^2 c^2)^{1/2}$ and the four-dimensional Pauli-Lubanski operator with components $\hat{W}_0 = (\hat{\boldsymbol{P}}\cdot\hat{\boldsymbol{J}})$, $\hat{\boldsymbol{W}} = c^{-1}\hat{H}\hat{\boldsymbol{J}} - c\hat{\boldsymbol{P}}\times\hat{\boldsymbol{K}}$. Then, the spin and the Newton-Wigner position operator for particles with mass are given by $\hat{S} = \hat{W}/\hat{M}c - \hat{W}_0 \hat{P}/[\hat{M}(\hat{M}c^2 + \hat{H})]$ and $\hat{R} = -(c^2/2)(\hat{H}^{-1}\hat{K} + \hat{K}\hat{H}^{-1}) - c^2 \hat{P} \times \hat{S}/[\hat{H}(\hat{M}c^2 + \hat{H})]$. For spinless particles, it is also possible to define wavefunctions in the momentum representation, $\psi(\boldsymbol{p})$, which transform under the action of the operators defined above as

$$\hat{P}_i \psi(\boldsymbol{p}) = p_i \psi(\boldsymbol{p}), \quad \hat{R}_i \psi(\boldsymbol{p}) = i\hbar \partial \psi(\boldsymbol{p})/\partial p_i, \quad \hat{H}\psi(\boldsymbol{p}) = \omega_{\boldsymbol{p}} \psi(\boldsymbol{p}), \quad (5)$$
$$\hat{K}_i \psi(\boldsymbol{p}) = i\hbar(-c^{-2}\omega_{\boldsymbol{p}} \partial/\partial p_i - p_i/2\omega_{\boldsymbol{p}})\psi(\boldsymbol{p}), \quad \hat{J}_i \psi(\boldsymbol{p}) = -i\hbar(p_j \partial/\partial p_k - p_k \partial/\partial p_j)\psi(\boldsymbol{p}),$$

with $\omega_{\boldsymbol{p}} = c(m^2 c^2 + p^2)^{1/2}$ and $m$ the eigenvalue of the mass operator; in the last formula above $i, j, k$ form a right-handed coordinate system. Similarly, the wavefunction in the position representation, $\varphi(\boldsymbol{r})$, transforms according to

$$\hat{R}_i \varphi(\boldsymbol{r}) = r_i \varphi(\boldsymbol{r}), \quad \hat{P}_i \varphi(\boldsymbol{r}) = -i\hbar \partial \varphi(\boldsymbol{r})/\partial r_i,$$
$$\hat{H}\varphi(\boldsymbol{r}) = c(m^2 c^2 - \hbar^2 \nabla^2)^{1/2} \varphi(\boldsymbol{r}), \quad \hat{J}_i \varphi(\boldsymbol{r}) = -i\hbar(r_j \partial/\partial r_k - r_k \partial/\partial r_j)\varphi(\boldsymbol{r}), \quad (6)$$
$$\hat{K}_i \varphi(\boldsymbol{r}) = (c/2)[(m^2 c^2 - \hbar^2 \nabla^2)^{1/2} r_i + r_i (m^2 c^2 - \hbar^2 \nabla^2)^{1/2}]\varphi(\boldsymbol{r}),$$

where $r_x \equiv x$, $r_y \equiv y$, $r_z \equiv z$.

## 3. RELATIVISTIC TRANSFORMATIONS IN PHASE SPACE

The relativistic wavefunctions defined in the previous section and their inertial transformations constitute the formal basis of the quantum special relativity, as developed in [4]. However, there are conceptual differences between special relativity and quantum mechanics that cannot be easily reconciled. The arguments include (i) the position operator of a quantum particle, which is not a good observable since its exact measurement results in large momentum and energy uncertainties that can be associated with particle-antiparticle pair creation, and (ii) the particle localization concept. A quantum particle cannot be invariantly localized, since different observers would not agree upon the localization of a particle, and, moreover, well-localized quantum states spread when evolving and there is a non-vanishing probability of superluminal propagation. Although all these apparent difficulties can be dealt with in the standard quantum mechanical framework (see for instance the arguments in [4]), a phase space formulation of quantum mechanics can help alleviate some of these complications.

More precisely, we choose as mathematical tool for PS characterization of quantum systems the Wigner distribution function (WDF) [1], defined as

$$W(\mathbf{r}, \mathbf{p}) = h^{-3} \int \varphi^*\left(\mathbf{r} - \frac{\mathbf{r}'}{2}\right) \varphi\left(\mathbf{r} + \frac{\mathbf{r}'}{2}\right) \exp(-i\mathbf{p}\mathbf{r}'/\hbar) d\mathbf{r}'$$
$$= h^{-3} \int \psi^*\left(\mathbf{p} - \frac{\mathbf{p}'}{2}\right) \psi\left(\mathbf{p} + \frac{\mathbf{p}'}{2}\right) \exp(i\mathbf{p}'\mathbf{r}/\hbar) d\mathbf{p}' \qquad (7)$$

where $\mathbf{pr}$ is a shorthand notation for $p_x r_x + p_y r_y + p_z r_z$. The properties of the WDF and its relations to other PS distributions can be found in [1-2,6-7]. We only remark here that it is defined on classical PS variables, that it can be regarded as a quasi-probability of finding a quantum particle in PS (it is not a true probability since, although real-valued, it can be negative in certain regions), and that it cannot be localized in PS regions whose projection area on any of the conjugate planes $(r_i, p_i)$ is smaller than $h/2$. Since this area is a canonical invariant, one can regard a quantum particle characterized by a WDF as a PS spatially extended particle, in contrast to the classical particle of the relativity theory, which is perfectly localized in PS. However, unlike in the Schrödinger formulation of quantum mechanics, the WDF remains localized during canonical evolution, i.e. the WDF does not spread: only its shape can change, but not its localization area. As such, argument (ii) is not an issue in the PS formulation.

Although equivalent to the standard Heisenberg and Schrödinger formulations [8], the PS formulation of quantum mechanics relies on commutative position and momentum coordinates being, however, not incompatible with the commutation relations between conjugate operators (see the demonstration in the Appendix). In fact, *r* and *p* in PS are not sharp eigenvalues associated with position and momentum operators, respectively, since eigenstates of such operators do not exist [8], but are coordinates of a smooth distribution function. As such, a state cannot be precisely localized in PS, and there is no conceptual clash with the Heisenberg uncertainty principle. On the other hand, an extended PS distribution function is not an extended classical particle: the WDF is an amplitude probability for the quantum particle and hence does not describe a mass distribution. The meaning of *r* and *p* in the PS formulation of quantum mechanics prevents any attempt to precisely measure the eigenvalues of these operators, invalidating argument (i) above.

The Poincare group approach to quantum special relativity in PS relies on the fact that, if the wavefunction transforms according to $\varphi'(r) = U(r, (\hbar/i)\nabla_r)\varphi(r)$, where $\nabla_r$ is the gradient operator in the *r* space, the corresponding WDF changes as [6]

$$W'(r, p) = \left[ U\left(r - \frac{\hbar}{2i}\nabla_p, p + \frac{\hbar}{2i}\nabla_r\right) - U\left(r + \frac{\hbar}{2i}\nabla_p, p - \frac{\hbar}{2i}\nabla_r\right) \right] W(r, p). \quad (8)$$

Therefore, the quantum wavefunction transformations from one frame of reference to another, as expressed in equations (5) and (6), can be easily translated in changes in the WDF. In the following we exemplify the effect of the relativistic inertial transformations on a quantum state with the wavefunction

$$\varphi(x) = (a^2/\pi)^{1/4} \exp(-a^2 x^2/2) \quad (9)$$

in the reference frame *F*, the corresponding WDF being given by

$$W(x, p) = (2/h)\exp(-a^2 x^2 - p^2/a^2\hbar^2). \tag{10}$$

The contour plots of this WDF are represented in Fig. 1 in normalized coordinates $X = ax$, $P = p/a\hbar$. We consider in this section only one-dimensional wavefunctions for graphical purposes and denote the momentum component with *p* for simplicity; it is not possible to represent the WDF of higher-dimensional wavefunctions. For one-dimensional wavefunctions it is not possible to apply an angular momentum operator as defined in the previous section, but we concentrate on the other inertial transformations. Our intention is to illustrate the effect of the inertial transformations on quantum wavefunctions in PS, and to compare the relativistic with the non-relativistic changes in the WDF. The WDF in the inertial frames $F'$ is denoted by $W'(x, p)$, the corresponding distribution function obtained in the non-relativistic (Galilei) approximation of the inertial transformations being labeled by the subscript *nr*. Through the analysis in this section we demonstrate that the WDF, unlike the quantum wavefunction, remains localized during inertial transformations, so that observers in different reference frames agree upon the localization (in the same, finite PS area) of quantum particles. The WDF does not spread, i.e. the PS area in which it is localized remains the same, although its form can change, so that superluminal propagation is not an issue. Quantum mechanics becomes thus compatible with special relativity, if the localization criterion is relaxed from a delta-like distribution in PS to a finite PS area.

To start with, if the wavefunction in (9) is subjected to a time translation *t*, the relativistic WDF becomes

$$W'(x,p) = \hbar^{-2}\pi^{-3/2}a^{-1}\exp(-p^2/a^2\hbar^2)$$
$$\times \int \exp[-p'^2/4a^2\hbar^2 + ip'x/\hbar + (itc/\hbar)(\sqrt{m^2c^2+(p+p'/2)^2} - \sqrt{m^2c^2+(p-p'/2)^2})]dp' \quad (11)$$

its contour plots being displayed in Fig. 2(a), whereas the non-relativistic WDF is

$$W'_{nr}(x,p) = (2/h)\exp[-a^2(x+tp/m)^2 - p^2/a^2\hbar^2], \quad (12)$$

and is represented in Fig. 2(b). In both simulations $m = 2a\hbar/c$, $t = 4/ac$. It is instructive to observe that in the non-relativistic limit the WDF suffers a shear transformation along $x$, preserving its form. This WDF transformation at time translations is known from the time evolution of free particles; in special relativity it just acquires a new meaning. Not known, however, is the relativistic PS transformation in Fig. 2(a), which is equivalent to an aberration in optics [9] because the WDF is distorted; relativistic time translations of quantum particles are not linear transformations.

A space translation with $x_0$ leads to a translation in the position coordinate of the WDF in both relativistic and non-relativistic cases:

$$W'(x,p) = W'_{nr}(x,p) = W(x-x_0, p). \quad (13)$$

The WDF after a space translation is represented in Fig. 3 for $x_0 = 1/a$. It is interesting to note that space translations of quantum particles are linear transformations, while time translations are not.

The WDF of a relativistic boost transformation is calculated as in (7) with a wavefunction in the momentum representation that changes as [4]

$$\psi'(p) = \left[\cosh\theta - \frac{p\sinh\theta}{\sqrt{m^2c^2+p^2}}\right]^{1/2} \psi(p\cosh\theta - \sqrt{m^2c^2+p^2}\sinh\theta) \, , \tag{14}$$

and is represented in Fig. 4(a) for $m = 2a\hbar/c$, $\theta = 0.15$ rad. In the non-relativistic limit of the boost the wavefunction transforms as $\psi'(p) = \psi(p-mv)$, the effect of the WDF being equivalent to a translation in the momentum coordinate:

$$W'_{nr}(x,p) = W(x, p-mv). \tag{15}$$

Such a translation is equivalent in optics with the action of an ideal lens, the relativistic transformation being similar with the effect of an aberrated lens [9]. Again, the relativistic transformation of the quantum wavefunction is not linear. However, the WDF distortion is distinctively different from the case of translation in time, different inertial transformations acting in a specific manner on the WDF.

As shown in the simulations above, a quantum particle remains localized in PS during inertial transformations; the WDF occupies the same PS area, as expected from its general properties. However, the WDF transformations from one reference frame to another have not been explicitly calculated until now. The distortion of the WDF at relativistic transformations does not come as a surprise, but the results in this paper illustrate the exact form of PS quantum distortions.

The importance of this first PS study of inertial transformations of a quantum wavefunction is to strengthen the connection between special relativity and quantum mechanics. Quantum particles do remain localized in PS when subject to inertial transformations, the only difference from classical particles being the finite area of localization in PS, which is determined by the Planck's constant $h$. The preservation of quantum localization concept for different reference frames holds only in PS.

It is interesting to note that the finite PS area associated to a quantum particle is not equivalent to the finite four-dimensional extent of an event introduced in [10]. We have not described the evolution of the quantum system in the Minkowsky four-dimensional space, but have used instead the Poincare group approach to relativistic transformations. Therefore, time is not the fourth coordinate of an event but a parameter of an inertial transformation generated by the Hamiltonian. This approach is more suitable for a relativistic theory of quantum systems, where the time parameter has a special and distinct meaning than spatial coordinates, while preserving the essence of relativistic transformations, i.e. invariance of the laws of physics with respect to inertial frames of references. The quantum wavefunction discontinuity propagates, in fact, according to classical laws [11].

CONCLUSIONS

We have developed a PS formulation of inertial transformations applied to a quantum system. In this formulation the PS area occupied by the quantum particle is invariant in different reference frames and thus quantum particles can be regarded as localized by different observers, although the localization is to be understood as occurring in a finite region of PS. The results of this paper indicate that the PS approach to special relativity of quantum systems has a certain advantage compared to usual quantum treatments because, unlike quantum wavefunctions, the WDF changes only its form but not its localization at inertial transformations. Moreover, the relativistic transformations differ in PS from non-relativistic transformations through the distortion of the WDF (with the exception of space translations), effect that is intuitively represented in the PS formulation. The distortions of the WDF are different for different transformations.

## APPENDIX
We show here that the Heisenberg uncertainty relation for a one-dimensional quantum wavefunction $\varphi(x)$, i.e. $\Delta x \cdot \Delta p \geq \hbar/2$, where $(\Delta x)^2 = \langle (x-\langle x \rangle)^2 \rangle = \langle x^2 \rangle - \langle x \rangle^2$, a similar relation existing for $\Delta p$, is compatible with the PS formulation of quantum mechanics; the generalization to higher-dimensional quantum wavefunctions is obvious. The Heisenberg uncertainty relation follows from the commutation relation $[\hat{x}, \hat{p}] = i\hbar$ between the position and momentum operators, the expectation value of any function of these operators, $f(\hat{x}, \hat{p}) = f(x, -i\hbar\partial/\partial x)$, being calculated as $\langle f(x, -i\hbar\partial/\partial x) \rangle = \int \Psi^*(x,t) f(x, -i\hbar\partial/\partial x) \Psi(x,t) dx$; when products of $x$ and $p$ are encountered, their order is not arbitrary.

Although both the commutation and uncertainty relations are defined in phase space, the corresponding commutator for the classical phase space coordinates vanishes, i.e. $[x, p] = 0$, signifying that the WDF is well defined at any point in PS. These statements are not contradictory since the uncertainties in $x$ and $p$ satisfy the Heisenberg inequality if the expectation value is understood as PS average. More precisely, if the PS expectation value of a function of $x$ and $p$ is defined as $\overline{f(x,p)} = \int f(x,p) W(x,p) dx dp$ for a normalized WDF, the average PS values for position and momentum become identical to the quantum expectation values for the operators $\hat{x} = x$ and $\hat{p} = -i\hbar\partial/\partial x$:

$$\bar{x} = \int \Psi^*\left(x - \frac{x'}{2}; t\right) x \Psi\left(x + \frac{x'}{2}; t\right) \exp(-ipx'/\hbar) dx' dx dp = \int \Psi^*(x;t) x \Psi(x;t) dx = \langle \hat{x} \rangle \quad \text{(A1)}$$

$$\bar{p} = \int \Psi^*\left(x - \frac{x'}{2}; t\right) p \Psi\left(x + \frac{x'}{2}; t\right) \exp(-ipx'/\hbar) dx' dx dp = \int \Psi^*(x;t) \left(-i\hbar\frac{\partial}{\partial x}\right) \Psi(x;t) dx = \langle \hat{p} \rangle$$
(A2)

Generally, $\langle f(\hat{x}, \hat{p}) \rangle = \overline{f(x,p)}$. Thus, the WDF reconciles quantum and classical physics, being compatible with the quantum commutation relation, although defined on the classical phase space.

FIGURE CAPTIONS

Fig. 1   The WDF of a Gaussian wavefunction.

Fig. 2   The effect of a (a) relativistic and (b) non-relativistic translation in time on the WDF in Fig. 1.

Fig. 3   The effect of a translation in space on the WDF in Fig. 1

Fig. 4   The effect of a (a) relativistic and (b) non-relativistic boost on the WDF in Fig. 1

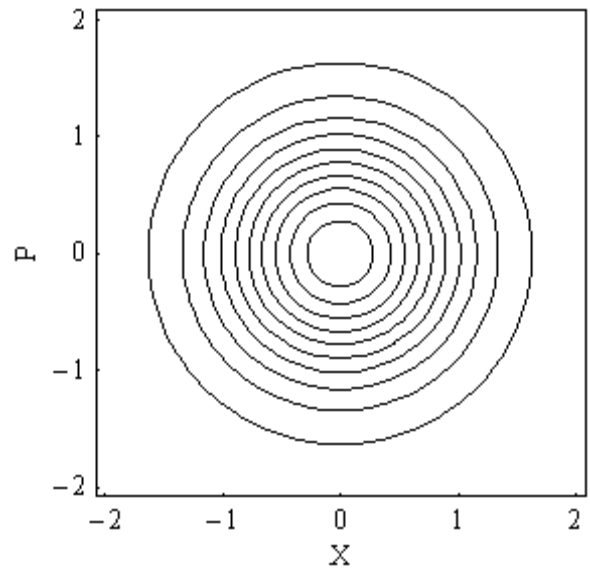

Fig. 1

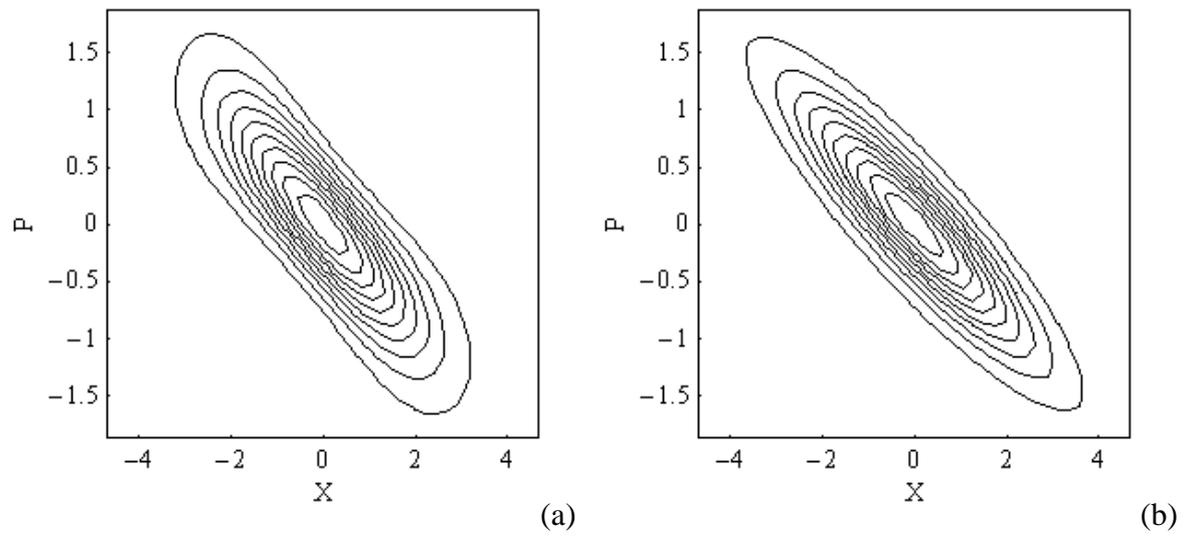

(a) (b)

Fig. 2

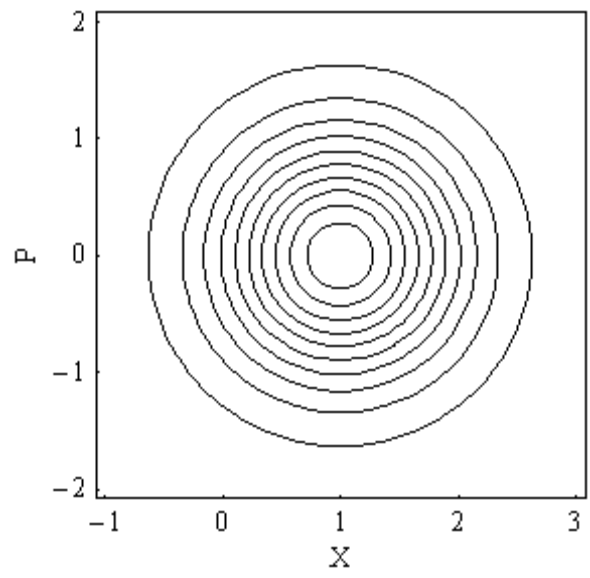

Fig. 3

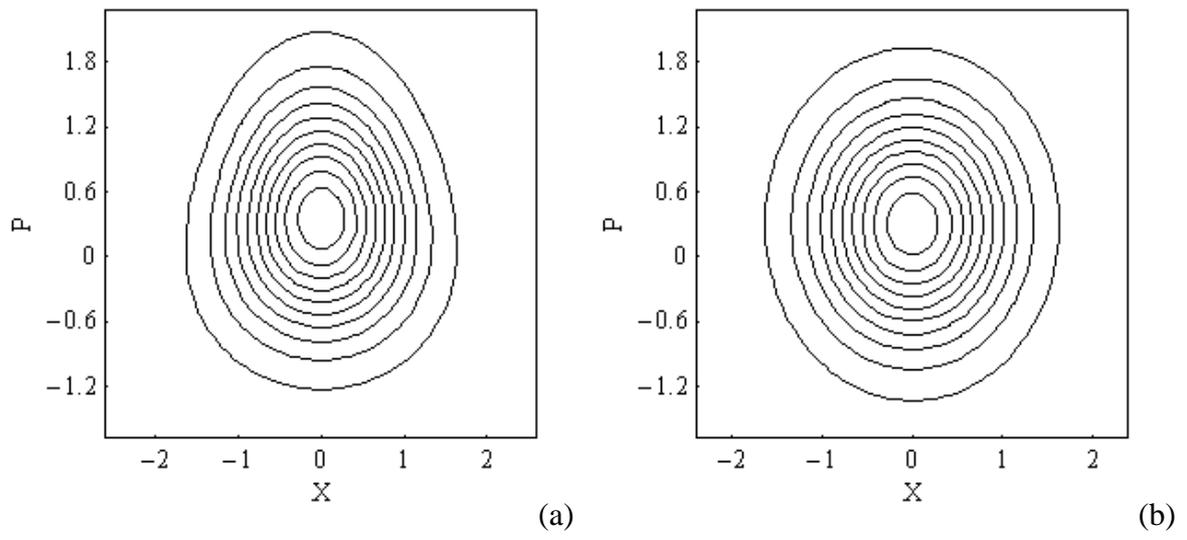

(a) (b)

Fig. 4